\documentclass{iaus}
\usepackage{graphics}
  \checkfont{eurm10}
  \iffontfound
    \IfFileExists{upmath.sty}
      {\typeout{^^JFound AMS Euler Roman fonts on the system,
                   using the 'upmath' package.^^J}%
       \usepackage{upmath}}
      {\typeout{^^JFound AMS Euler Roman fonts on the system, but you
                   dont seem to have the}%
       \typeout{'upmath' package installed. iaus.cls can take advantage
                 of these fonts,^^Jif you use 'upmath' package.^^J}%
      }
  \else
  \fi

  \checkfont{msam10}
  \iffontfound
    \IfFileExists{amssymb.sty}
      {\typeout{^^JFound AMS Symbol fonts on the system, using the
                'amssymb' package.^^J}%
       \usepackage{amssymb}%

      }{}
  \fi

  \IfFileExists{amsbsy.sty}
    {\typeout{^^JFound the 'amsbsy' package on the system, using it.^^J}%
     \usepackage{amsbsy}}
    {}

\newsavebox{\astrutbox}
\sbox{\astrutbox}{\rule[-5pt]{0pt}{20pt}}

\title[The Interplay among Black Holes, Stars and ISM in Galactic 
       Nuclei]{The parsec scale region of Active Galactic Nuclei in the IR}

\author[M. A. Prieto]%
{M. Almudena Prieto, Klaus Meisenheimer}

\affiliation{Max-Planck Institute for Astronomy, Heidelberg, Germany 
email: prieto@mpia.de\\[\affilskip]
}

\pubyear{2004}
\volume{222}

\pagerange{1--8}
\date{?? and in revised form ??}
\jname{The Interplay among Black Holes, Stars and ISM \\in Galactic Nuclei}
\editors{Th. Storchi Bergmann, L.C. Ho \& H.R. Schmitt, eds.}
\begin{document}

\maketitle

\begin{abstract}

First results from the AGN-Heidelberg program aimed at spatially
resolving the central pc region of the closest Active Galactic Nuclei
are presented. The core region of prototype active nuclei are clearly
unveiled at IR waves and at distances from the nucleus - few pc- where
circumnuclear starforming regions appear not to be present.  Within
that perspective, classical active nuclei as Circinus and NGC 1097,
reveal with unprecedented detail clear channels of material being
driven to the core whereas others as Centaurus A and NGC 1566, show a
"clean" core environment. At the very center, a central compact region
of about 2 pc scale is resolved in Circinus but not in the other cases
challenging thus the universal presence of the putative obscuring
torus.

\end{abstract}

\firstsection % if your document starts with a section,
              % remove some space above using this command.
\section{Introduction}

The Active Galactic Nuclei (AGN) paradigm relay on a massive black hole
surrounded by an accretion disk with size of few astronomical
units. Surrounding this core, there is the broad line region, 
still spatially unresolved  with a predicted size of less than 1 pc. 
To account for the diversity of AGN, a further additional component is
introduced: a central obscuring structure - a torus/warp-disk shape - which
prevents photons from the central engine to escape  in an
isotropic manner. The presence of collimated nuclear cones of ionized
gas constrains the size of this obscuring structure to be between 1 to  
a few tents of a pc. If due to dust,  the  peak emission of 
this structure should be in
the IR and therefore observational confirmation of it  
has  been hampered   due to spatial resolution limitations in this
range. Diffraction limited observations in 8-10m class telescopes
and long-base-line interferometry  allows us for the first time to 
resolve this structure  in
the IR for at least the closest AGN. This is the goal of the AGN-Heidelberg program.  This paper  presents  
sub-arcsec resolution observations in the near IR
conducted with the VLT of the closest and brightest AGN accessible from the
South Hemisphere.  The observations allows us to set very
stringent constrains on the size of the AGN core and hence, on
that of the torus. 

\section{Target Sample and observations}
The  sample of targets studied in this program 
includes those Southern objects having 
 10um flux larger than 300 mJy, a total of 18 targets.
This brightness criteria was introduced to guaranty follow up
observations with VLTI/MIDI  interferometry at 10um for the most suitable
cases.   Broad-band J to M images for the sample are  being collected 
 with the
Adaptive-Optics-assisted NACO IR camera/spectrograph   at the  VLT. 
The best spatial  resolution is being achieved in the Ks-band,
with typical FWHM:  $0.07'' < FWHM <0.16''$. 
At the distance of  the targets  presented here, 
this  is equivalent to scales of 1 to 10 pc.

\section{First results: the nearest AGN}

A summary of the results obtained for for the four
more relevant AGN studied:   Circinus,
Centaurus A,  NGC 1097 and NGC 5506 are discussed bellow.

\begin{figure}
\centering
\resizebox{6.5cm}{!}{\includegraphics{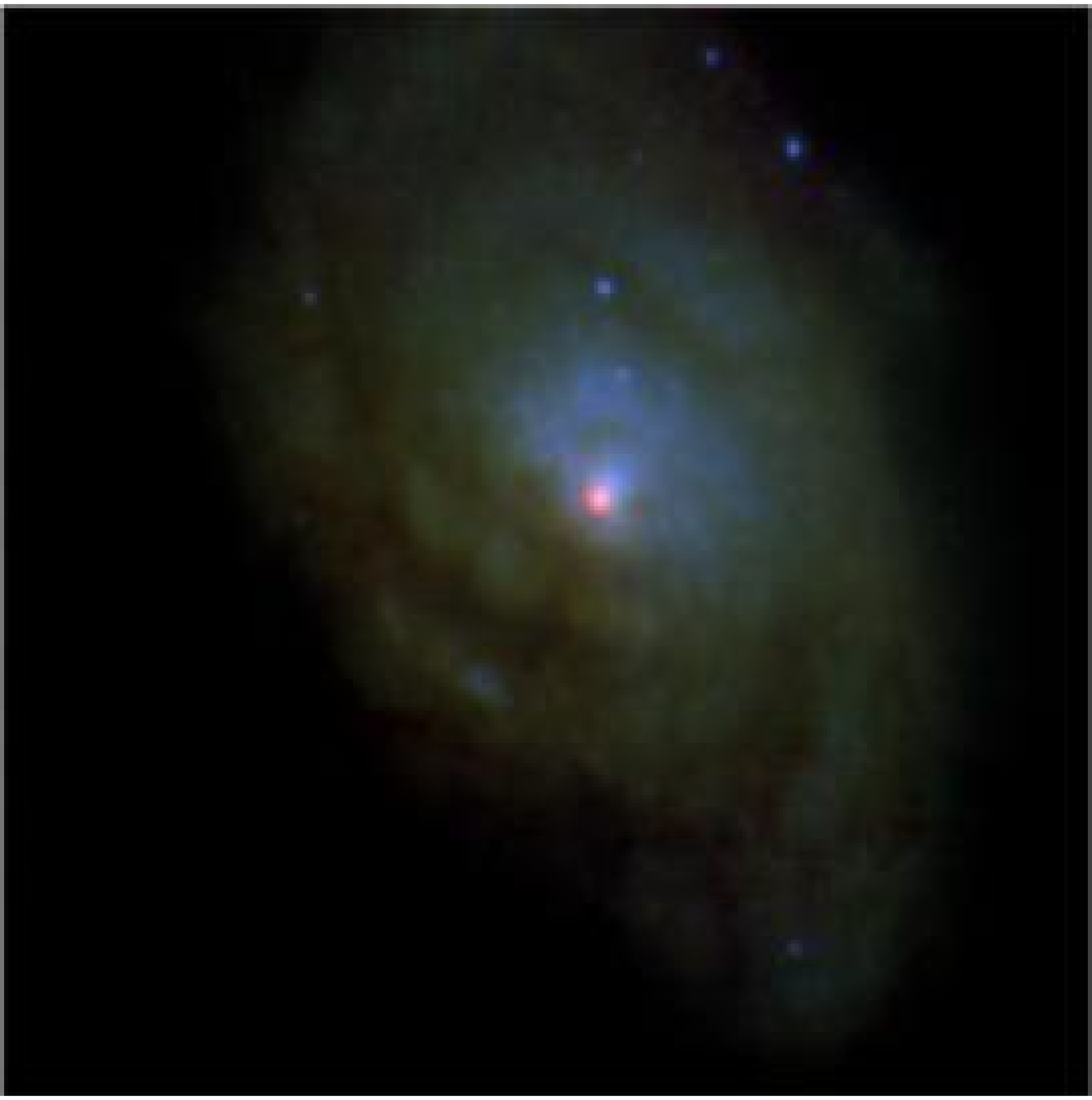}}
\resizebox{6.5cm}{!}{\includegraphics{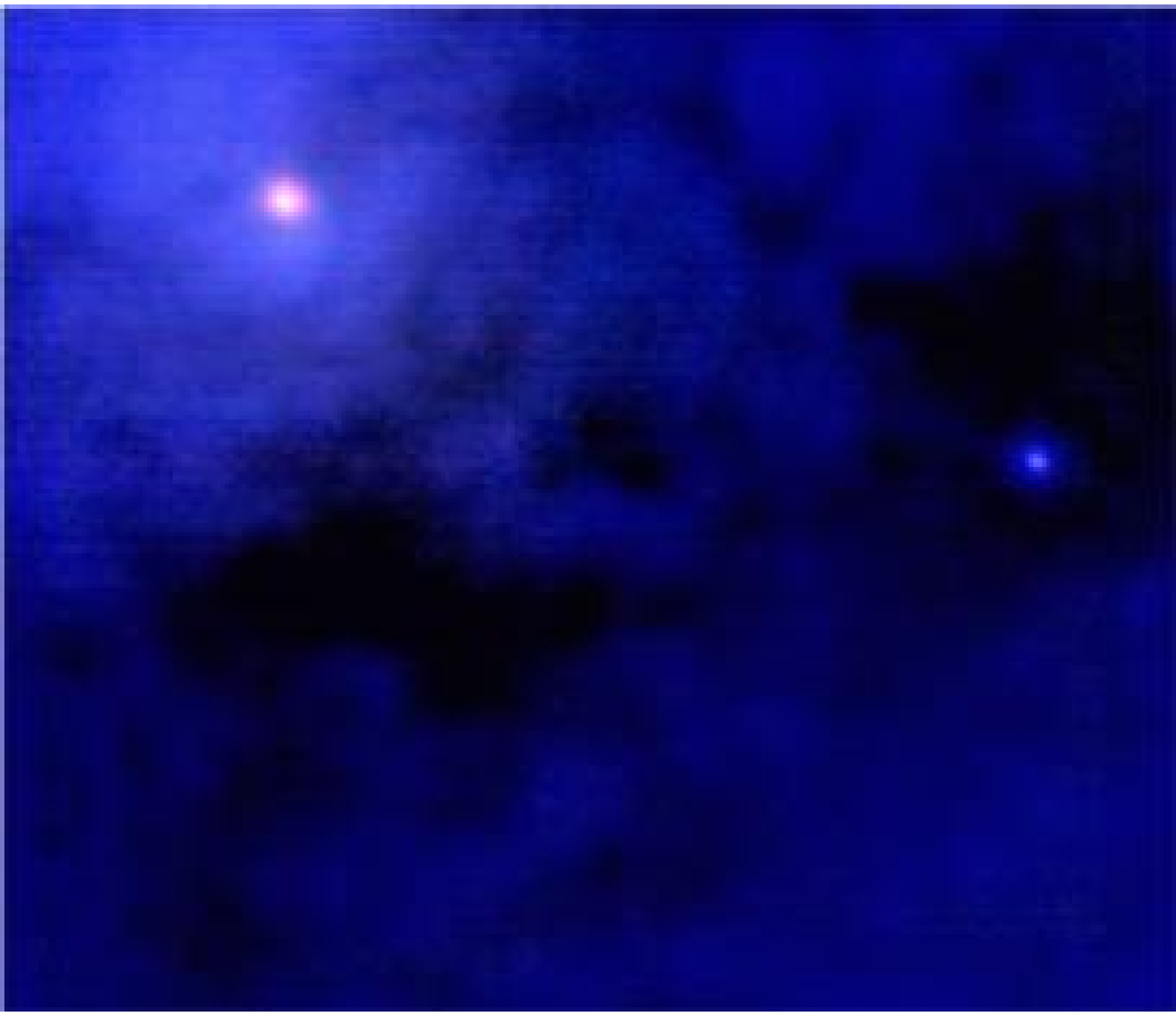} }
\caption[]{Left: True
 colour image of Circinus,  FoV = 27''x27'',
 combining HST F814W (blue) and NACO J- (green) and Ks-bands
(red). Some of the stars used for astrometry and spatial resolution
estimate are visible.  Right: NACO J+H+K image of Cena A,
FoV=7''x14''.  The star on the right side was used for astrometry and
spatial resolution estimate. North up, East left in both figures.}
\end{figure}

\subsection{Circinus}

Circinus is a SAb galaxy inclined by $\sim \rm65 \deg$. 1 arcsec
$\sim$ 19 pc. Several dust lanes hide a large fraction of its East
side, including its nucleus.  It shows a well known, one side
nuclear ionization cone extending on the kpc scale in the North-West
direction.  Circinus is the only case so far where we find a resolved
central core from K- to M- bands with a size FWHM= 1.9 $\pm$ 0.6 pc.
This size could be measured accurately due to the presence of several
stars in the same field.  A composite HST 8000 A and NACO J + K 
image is shown in Fig. 1. The core - red central point in the figure -
is only contributing in the Ks-band, and it is shifted by $ 0.15$
arcsecs (2.8 pc) Shout-East from the HST and NACO J-band peaks
emission.  It defines the vortex position of a rather collimate beam -
bluish central emission in the figure pointing North-West - with
extension $\sim$ 10 pc and seen both in continuum light below
1.6\,$\mu$m and in $H\alpha$ line. This colimated beam is  in
the direction of Circinus ionization cone.  The red core is also the
center position of a [SiVII]2.48um ionization ``double cone''
discovered also in our NACO observations.  The extinction in the
inmediate surroundings of the core, derived from colour maps after
comparing with those of normal ellipticals/spirals (e.g. Giovanardi \&
Hunt 1996), leads to Av=6 (screen dust layer) or Av=20 (gas mixed
dust).  Extensive analysis of these observations are presented in
Prieto et al.  (submitted to ApJ).

\subsection{Centauros A}

Together with Circinus, these are the two nearest AGN in the Southern
Hemisphere.  Cen A is the nearest  radio galaxy with a Seyfert type 2 nucleus,
 1 arcsec  is $\sim$16pc.  Because of the prominent
dust lanes covering Cen A, its nucleus is fully obscured at optical
wavelengths but it was unveiled in the IR by HST (Marconi et
al. 1999).  VLT provides 3 times resolution better and a NACO Ks-band
diffraction-limited image of Cen A sets a stringent upper limit for
the size of its core to be FWHM$<$1 pc (Haering et al. 2003). A NACO
J+H+K image of the central region (Fig. 1, right) shows a bright and
isolated core source sitting on top of rather diffuse emission from
the galaxy.  The extinction measured from J-H colors at the
surroundings of the core leads to values of Av$\sim$ 7 when comparing
them with those of normal bulges, and ssuming a foreground dust layer.

\subsection{NCC 1097}

This is one of the few nearest Seyfert 1 objects in the South: 1
arcsec $\sim$ 70 pc. The galaxy is a SBb with a prominent
circumnuclear ring at radius of 1.3 kpc enclosing a nuclear bar.  Fig 2
shows a NACO Ks-band image showing all those features.
 At the resolution of this  image, FWHM$\sim 0.14''\times 0.19''$,
the core of NGC1097 is unresolved which sets an upper limit to its
size of $\sim$ 11 pc.  The best detailed view of the core region is
however seen in the NACO J-K image (Fig. 2, right). It shows
 a complex network of
filaments of dust and gas spiraling down towards the core. Some of the
longest filaments seem to connect with the circumnuclear star forming
ring. Presumably, material from the star forming regions is loosing
angular momentum and falling straight to center to feed the AGN.
The extinction derived from colour maps, after comparing with those of
normal bulges lead to rather moderate values: Av=1.4 - 4, depending of
whether a screen case or gas mixed with stars case are respectively
considered (Prieto et al. in preparation).

\begin{figure}
\centering
\resizebox{6.5cm}{!}{\includegraphics{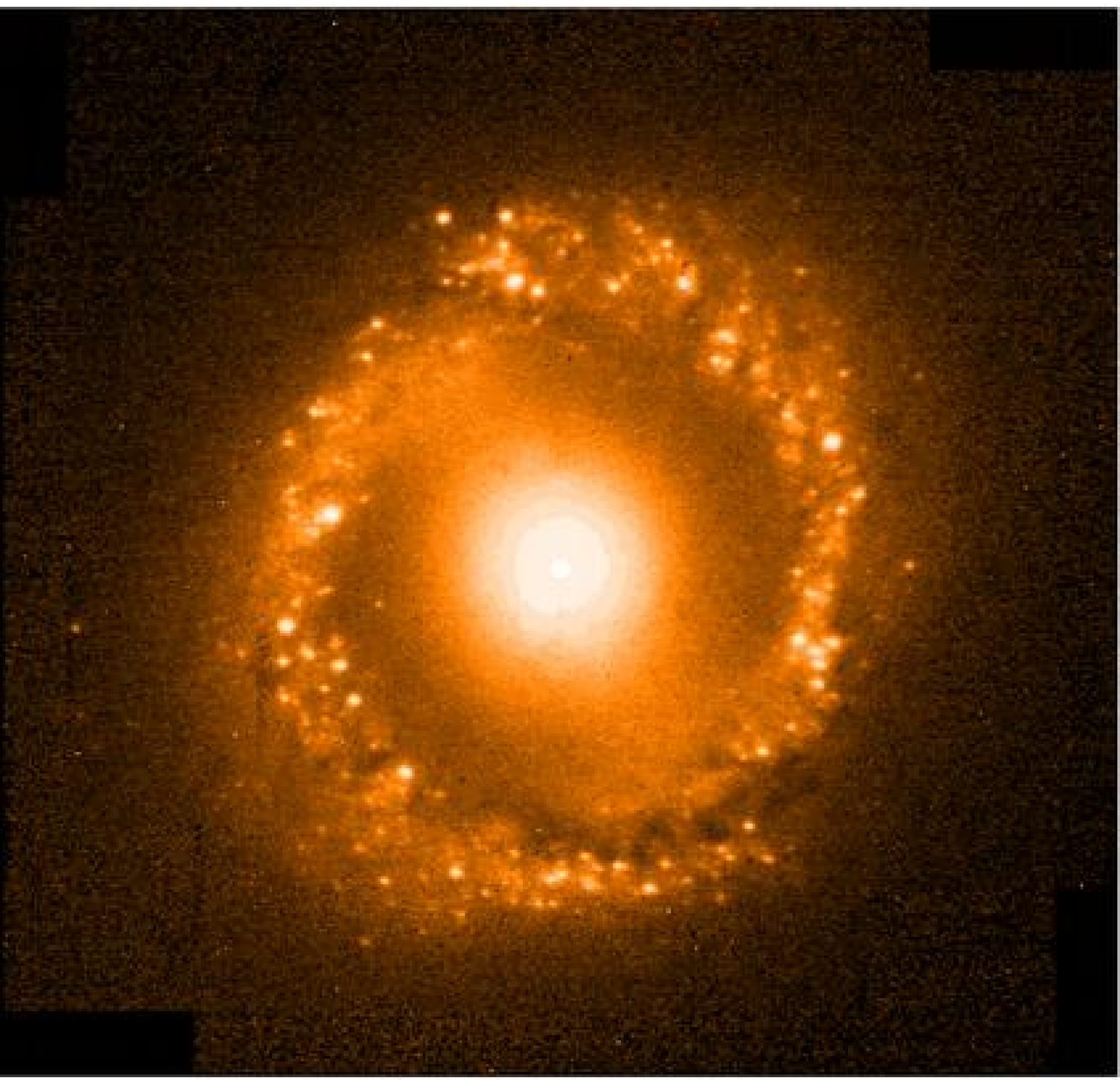}}
\resizebox{5cm}{!}{\includegraphics{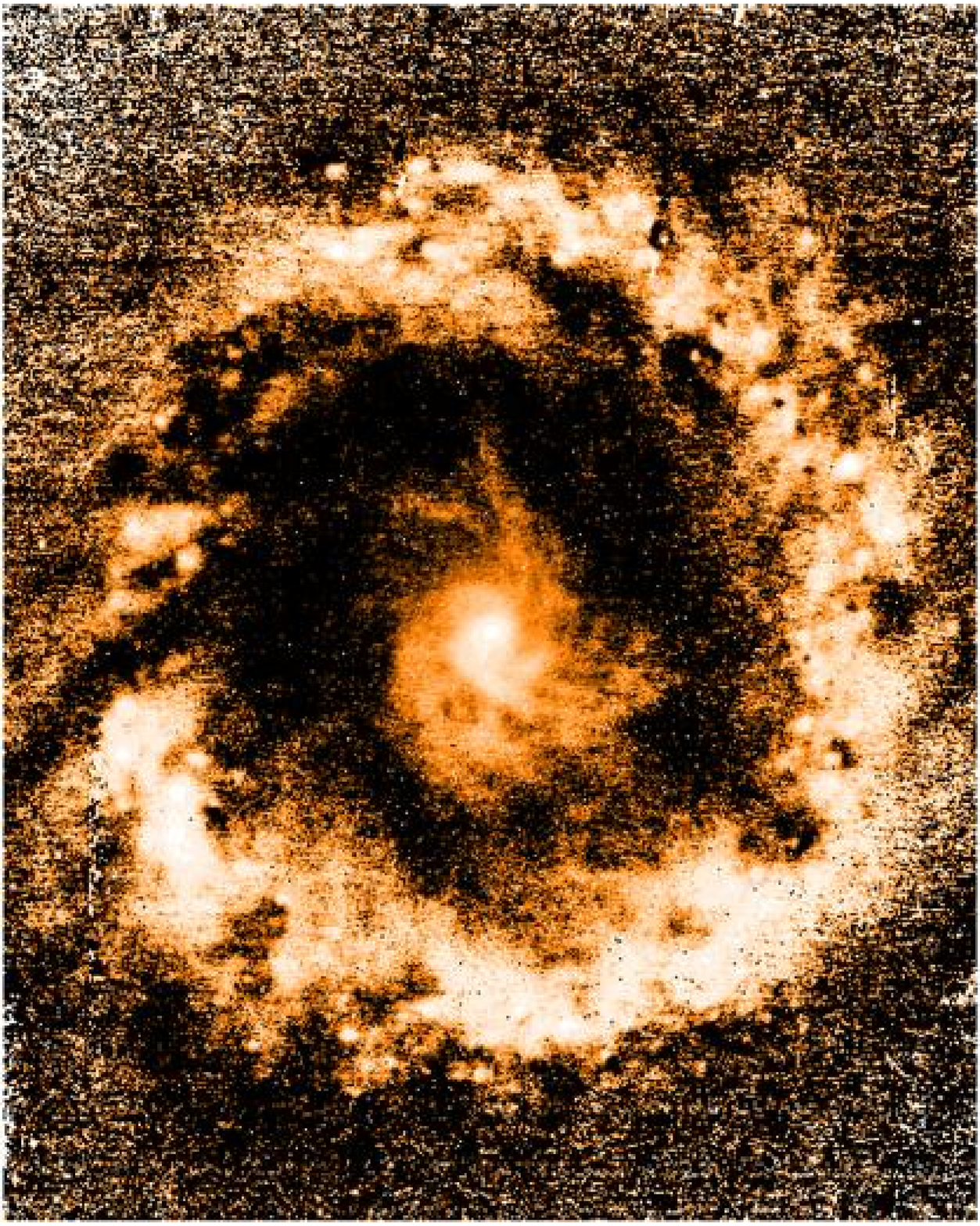} }
\caption[]{ NGC1097 IR view of the central 27x27  arcsec region, North is up, East to the left:  NACO Ks-band (left);  NACO J-K color image disclosing the cenral spiral network of dust and gas (right) .}
\end{figure}

\subsection{NGC 5506}

This is a disky,   edge on ({\it i} =70deg) galaxy covered 
by a thick dust lane all across its disk. Ussually refered as a
Seyfert type 2, the detection of broad both Pa$\beta$ and permitted OI
1.1287um line (Nagar et al. 2002) upgrades it into the  type 1 class.  
Fig. 3 (left panel) shows a HST 6000A
broad-band image of the central 35x35 arcsec  showing the dust lanes and the potential position of its nucleus.   NACO J- to L- band images
unveal a very bright  core but do not recover much from the
host galaxy. The L- and M-bands are limited in sensitivity but the Ks
image is deep to surface brightness $S_B(K)\sim <$16.  The Ks-band
image is diffraction-limited (Fig. 4, right panel), which sets un
upper limit to the size of the core of FWHM$<$10 pc.  J-K colors
measured around the core leads to Av$\sim >10$ when compared to those
of normal ellipticals/spirals, for which we assume J-K=0.95.  An
screen dust case is assumed.

\begin{figure}
\centering
\resizebox{6.5cm}{!}{\includegraphics{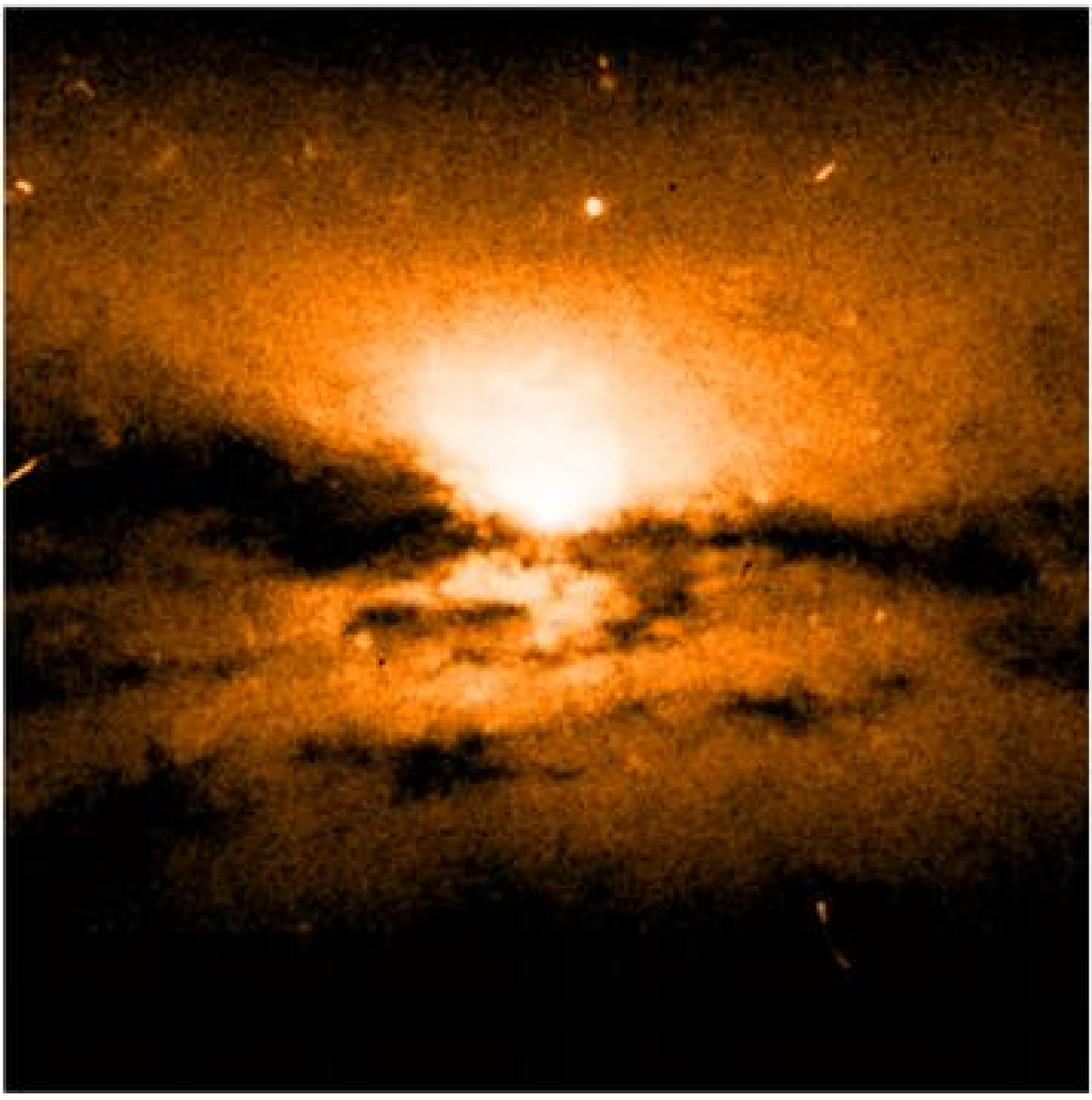}}
\resizebox{6.5cm}{!}{\includegraphics{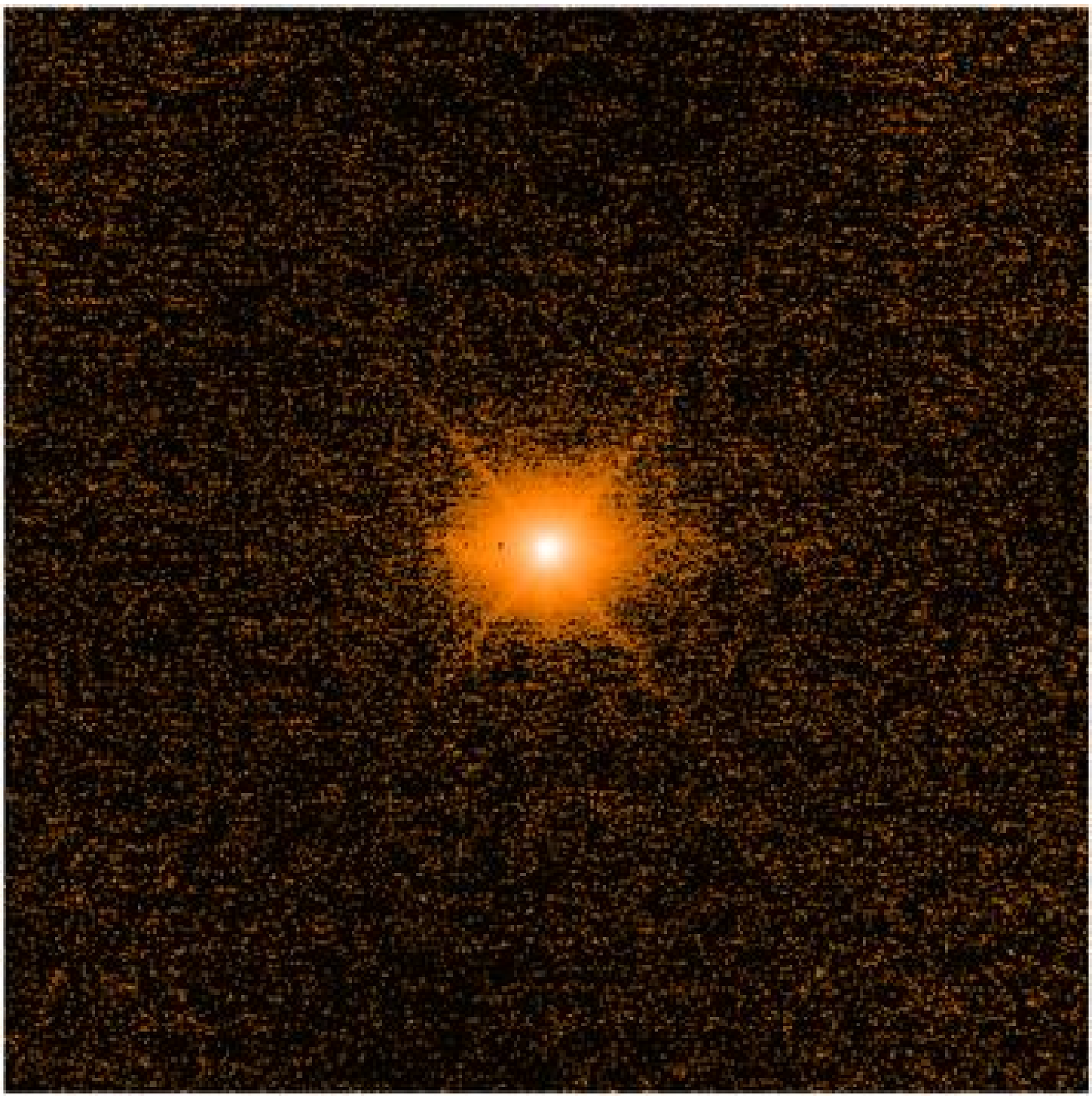} }
\caption[]{ NGC 5506 images of the central 35x45'' by  HST F606W (left) 
and  
NACO in  Ks-band, 20x20'' (right). Resolution of the NACO image is 57 mas. 
North is up, East, left.}
\end{figure}

\section{Conclusio: where is the torus?}

Considering the upper limit size of the core for the four galaxies
described, any torus in these objects has to be less than 10 pc in
N1097 and N5506 and definitively less than 1 pc in Cen A. As pointed
out to me by Dave Axon in this workshop, HST measures in Cen A a polarised
K-band nucleus (P= 11\%) from a size less than 1 pc (Capetti et
al. 2000).  If the polarization is due to scattered light from an
obscured nucleus, the size of the obscuring structure should be
strictly less than 1 pc. Of course, much cooler material, not traced
by these J to M- band observations, may extend to larger radius.
Indeed, as pointed out by Frank Israel, that may well be the case
considering the location of the HI absorption seen in the direction of
Cen A nucleus (Sarma et al. 2002).

Circinus is the only case where a central resolved structure is measured, 
 with a core size of FWHM$\sim$ 2pc. The J-to-N SED
of this core is compatible with dust temperature of about 300 K
(Prieto et al. 2004).  Coincidentally, these are about the size and
temperature measured in N1068 from the VLTI 10um spectrum (W. Jaffe,
this workshop).

\begin{acknowledgments}
I thank my collaborators  Juha Reunanem, Olivier Marco, 
Nadine Haering and Konrad Tristram.
\end{acknowledgments}

\end{document}